\documentstyle[manuscript,aps,prc,amssymb,tighten,graphicx]{revtex}
\begin{document}
\def\dsd{${^3S_1}$-${^3D_1}$\ }
\def\de{\Delta}
\def\kv{\bbox{k}}
\def\kvp{\bbox{k}'}
\def\kvpp{\bbox{k}''}
\draft
\title{Short-range effects on nuclear pairing}

\author{
W. Zuo$^{1}$, U. Lombardo$^{2,3}$, H.-J. Schulze$^{4}$, and C. W. Shen$^{3}$}

\address{
$^{1}$ Institute of Modern Physics, Lanzhou, China \\
$^{2}$ Dipartimento di Fisica, Universit\`a di Catania,
       57 Corso Italia, I-95129 Catania, Italy \\
$^{3}$ Laboratori Nazionali del Sud,
       Via Santa Sofia 44, I-95123 Catania, Italy \\
$^{4}$ INFN, Sezione di Catania,
       57 Corso Italia, I-95129 Catania, Italy }

\maketitle

\begin{abstract}
We determine the influence of the three-body force and the
medium modification of meson masses 
on pairing in nuclear and neutron matter.
A reduction of the pairing gap is found and increasing with density.
\end{abstract}



The pairing interaction in electron superconductors has a natural cutoff
determined by the lattice spacing.
On the contrary, in nuclear matter there is no such sharp cutoff.
The realistic bare nucleon-nucleon ($NN$) interactions have a smooth behavior
which is determined by the fit of the experimental phase shifts of high-energy
nucleon-nucleon scattering.
However, medium effects strongly modify the $NN$ interaction in nuclear and
neutron matter, including its short-range part \cite{rev}.
We have still a poor knowledge of these modifications.
All predictions on the pairing gap in nuclear matter suffer from this drawback,
as the solution of the gap equation is very sensitive to the value of the
tail of the interaction in momentum space \cite{umb},
i.e., to the short-range behavior of the coordinate-space potential.

In this note we discuss the role played in neutron and nuclear matter 
superfluidity by some mechanisms which may affect the short-range behavior 
of the pairing force.
The nucleon-nucleon correlations (ladder diagrams), which renormalize the
short-range part of the interaction much the same as in the Brueckner
$G$-matrix, fix already a kind of cutoff in the gap equation \cite{umb}.
Other processes may have a strong influence, such as nucleonic excitations
or $\overline{N}N$ exchange giving rise to
three-body forces (TBF) \cite{math}.
TBF have been proven to play a crucial role for the saturation properties of
nuclear matter due to their short-range repulsive nature \cite{lej}.
In addition, a medium modification of the heavy meson masses, which has been
addressed as manifestation of the chiral symmetry restoration in nuclear
matter\cite{bro}, can modify the range of the interaction.
Experimental evidence for the latter is provided by
dilepton production in high-energy heavy ion collisions \cite{cer}.

The magnitude of the pairing gap in nuclear matter is determined by the
competition between the repulsive short-range part and the attractive
long-range part of the interaction.
The medium modification of the long-range part was confidently obtained
within the RPA approximation or induced interaction model, and it was
discussed elsewhere \cite{rev,cla,schul}.
The short-range part is partially incorporated by the gap equation itself
with the bare interaction.
A simple way \cite{umb} to illustrate this property is to split the gap
equation into two coupled equations,
\begin{mathletters}\label{e:1}
\begin{eqnarray}
 \de_{\kv} &=& -\sum_{k'\le k_c}
 \widetilde{V}(\kv,\kvp) {\de_{\kvp} \over 2E_{\kvp} } \:,
\\
 \widetilde{V}(\kv,\kvp) &=&  V(\kv,\kvp) - \sum_{k'' \geq k_c}
 { V(\kv,\kvpp) \widetilde{V}(\kvpp,\kvp) \over 2E_{\kvpp} } \:,
\end{eqnarray}
\end{mathletters}
where
$E_{\kv} = \sqrt{(e_{\kv}-\mu)^2 + \de_{\kv}^2}$.
This shows that the effective interaction $\widetilde V$, arising from the
introduction of a cutoff $k_c$ in momentum space, sums up a series of ladder
diagrams analogous to the Bethe-Goldstone equation.
We want to stress that $\widetilde V$ and the cutoff are interrelated,
so that one cannot use a phenomenological interaction in the gap equation
and fix arbitrarily the cutoff.
It is quite sensitive to the tail ($k > k_c$)
of the $NN$ interaction, which reflects the
short-range part of the nuclear force.

\begin{figure}
\includegraphics[bb=120 710 490 710,angle=90,scale=0.67]{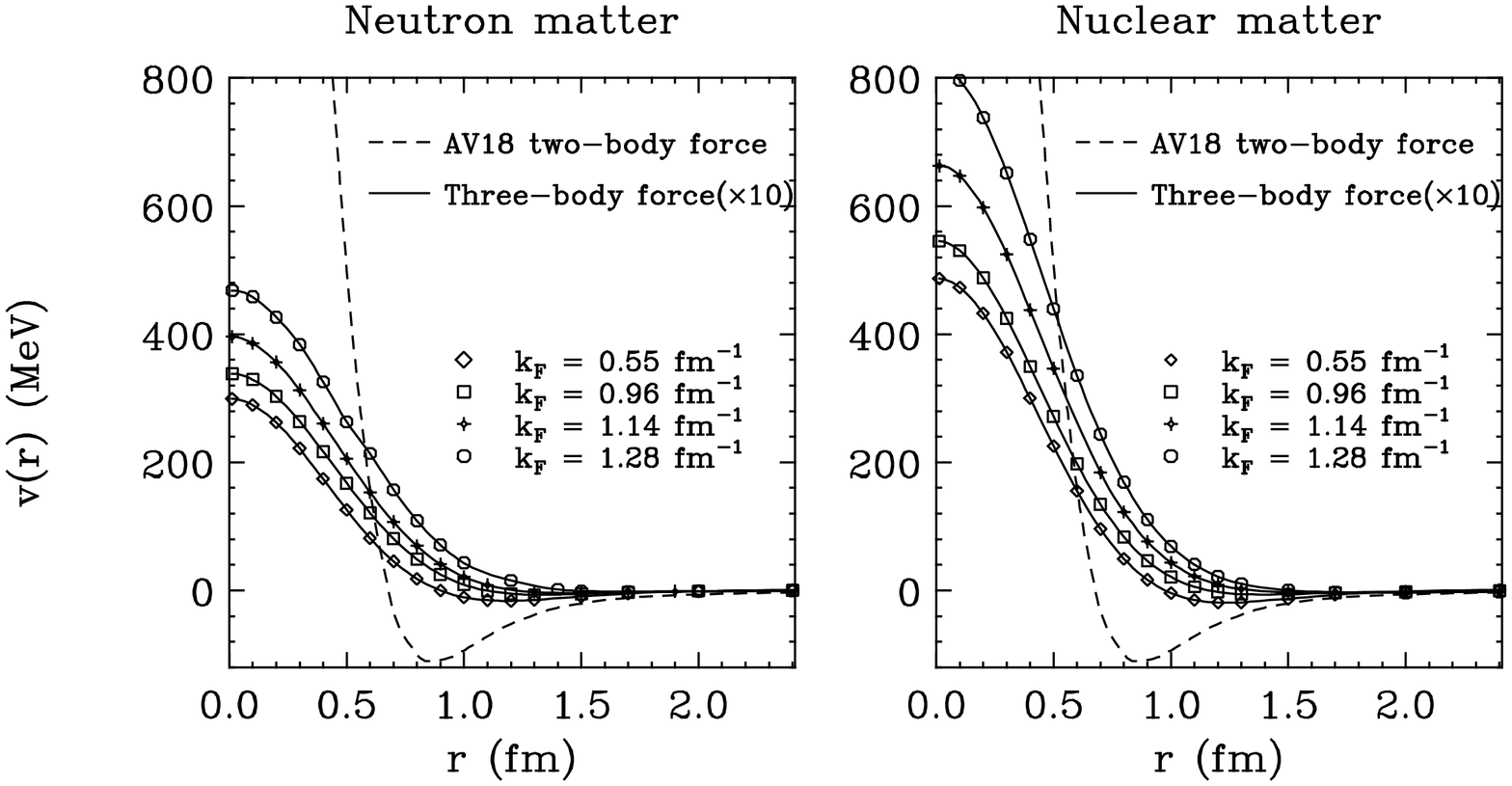}
\caption{
Nucleon-nucleon interactions vs.~distance.
The dashed curve shows the Argonne $V_{18}$ two-body force.
Solid curves with different symbols display the effective three-body force
(scale enlarged by a factor 10)
in neutron matter (left panel) and symmetric nuclear matter (right panel) 
at different neutron Fermi momenta, as indicated.}
\label{f:vr}
\end{figure}

A great deal of uncertainty is associated with the modification of the
short-range part of the two-body interaction when going from the vacuum to
the nuclear medium.
It is well know that the empirical saturation properties of nuclear matter
cannot be reproduced with the bare two-body force only, but that one has to
include also three-body forces which take into account the medium modification
of the in-vacuum $NN$ interaction \cite{lej}.
The effect of the three-body forces around the saturation region is mainly
to shift to lower density the balance between short-range repulsive and
long-range attractive components of the interaction, which mostly dominate
above the saturation density ($\rho_0 \approx 0.15-0.17\;\rm fm^{-3}$).
In the pairing problem we thus do not expect a remarkable effect in the
peak domain of the energy gap ($\rho\approx 0.02\;\rm fm^{-3}$).
In Fig.~\ref{f:vr} we plot the effective two-body force in the $^1S_0$ channel,
obtained from a meson-exchange model of the three-body force \cite{math}.
The density dependence arises from the averaging procedure adopted to reduce
the three-body force to an effective two-body force \cite{lej}.
In neutron matter the strength of $V_3(r)$ is smaller than in nuclear matter
due to the lack of the $T=0$ tensor force.
In either case the strength of $V_3(r)$ is very small in comparison to the 
two-body force plotted also for comparison in Fig.~\ref{f:vr}.
However, it is mainly effective on the short-range side
($r \lesssim 1\;\rm fm$)
and hence it affects $\widetilde V$.
The repulsion is increasing with rising density.

\begin{figure}
\includegraphics[bb=130 740 490 740,angle=90,scale=0.67]{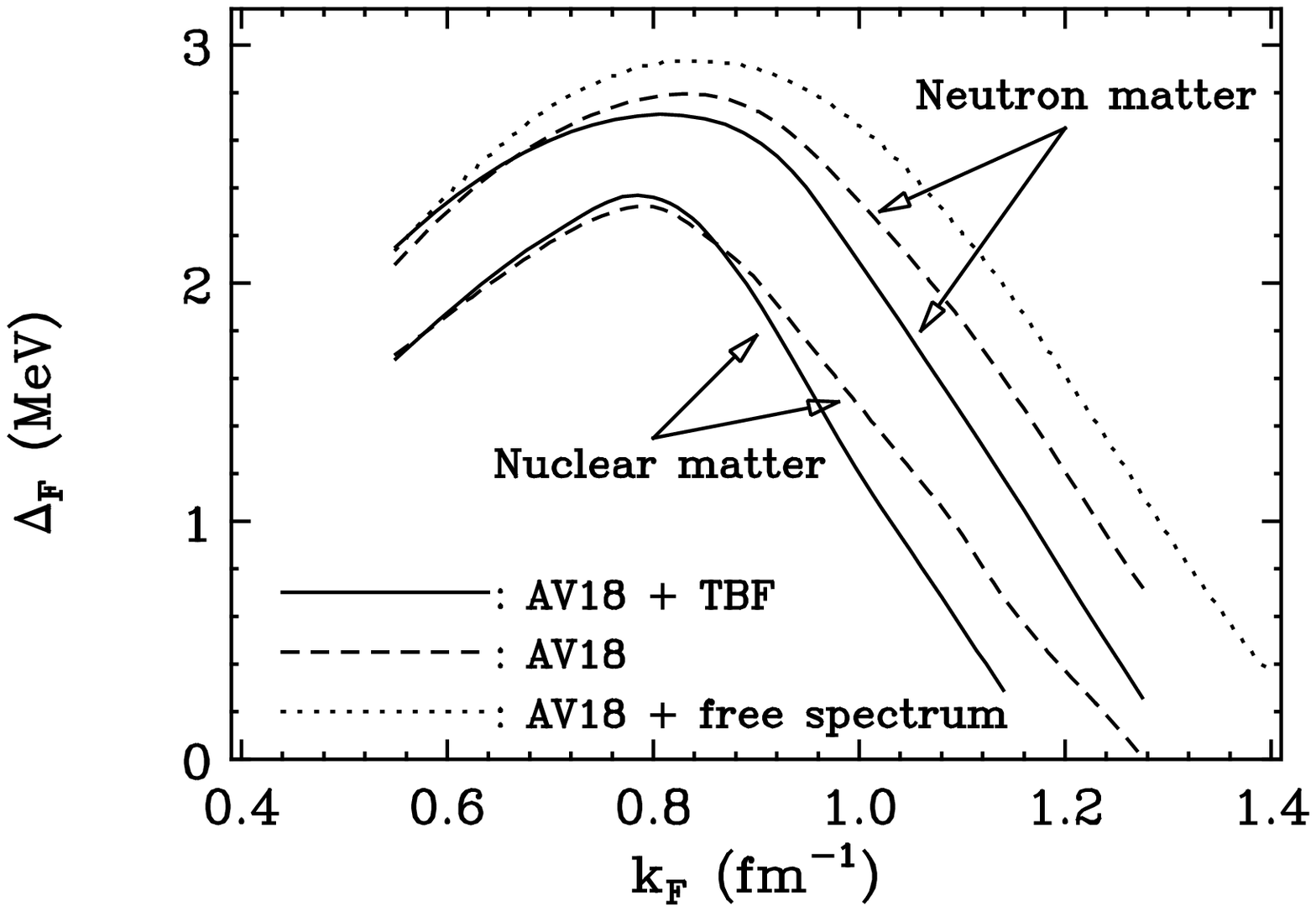}
\caption{
$^1S_0$ gap in neutron matter and in symmetric nuclear matter.
Solid curves show the results obtained using $AV_{18}$ plus three-body force.
Dashed curves are calculated adopting the pure $AV_{18}$ two-body force.
The dotted curve indicates the result using a free single-particle spectrum
and the bare $AV_{18}$ force.}
\label{f:gaptbf}
\end{figure}

In order to numerically investigate the effect of the three-body force we
have solved the gap equation, Eq.~(\ref{e:1}),
adding to the bare two-body force
(the Argonne $V_{18}$ potential\cite{wir})
the above three-body force.
At the same time the single-particle spectrum $e_{\kv}$ appearing
in the gap equation is computed using the same three-body force.
We have considered the two cases of $^1S_0$ pairing between like nucleons
embedded in neutron matter and nuclear matter.
The results for the energy gap $\de_F=\de(k_F)$ are
depicted in Fig.~\ref{f:gaptbf}.
The gap in nuclear matter is always smaller than in neutron matter,
since the single-particle spectrum is different in the two cases \cite{cug}.
As expected, the effect is a slight suppression of the gap,
quasi negligible at low density and increasingly larger
at higher density in both cases.

Since the masses of the exchanged mesons drive the range of the interaction,
it is also of interest to discuss in this note to which extent the energy
gap is influenced by the change of the properties of the
heavy mesons which act at short distances.
Unfortunately, at the present time this approach suffers a great deal
of uncertainty, since there is no clear idea about the medium modification of
the meson parameters.
The latter is motivated as being due to the partial restoration of chiral
symmetry, which causes hadron mass scaling \cite{bro} in the medium.
It appears experimentally supported by the quenching of the
$\rho$ meson mass in dilepton production in heavy ion collisions \cite{cer}.
The theoretical treatment \cite{san} can only be accomplished using an
interaction explicitly dependent on the meson parameters.
For this reason we use the Bonn-B potential \cite{mac} and only investigate
the effect of the $\omega$ meson mass,
whose strength is by far the largest one.
The Bonn-B potential has also been used recently in the relativistic theory
of pairing, where it predicts a gap comparable with the result of
the nonrelativistic BCS theory \cite{ring}.

\begin{figure}
\includegraphics[bb=120 690 490 690,angle=90,scale=0.73]{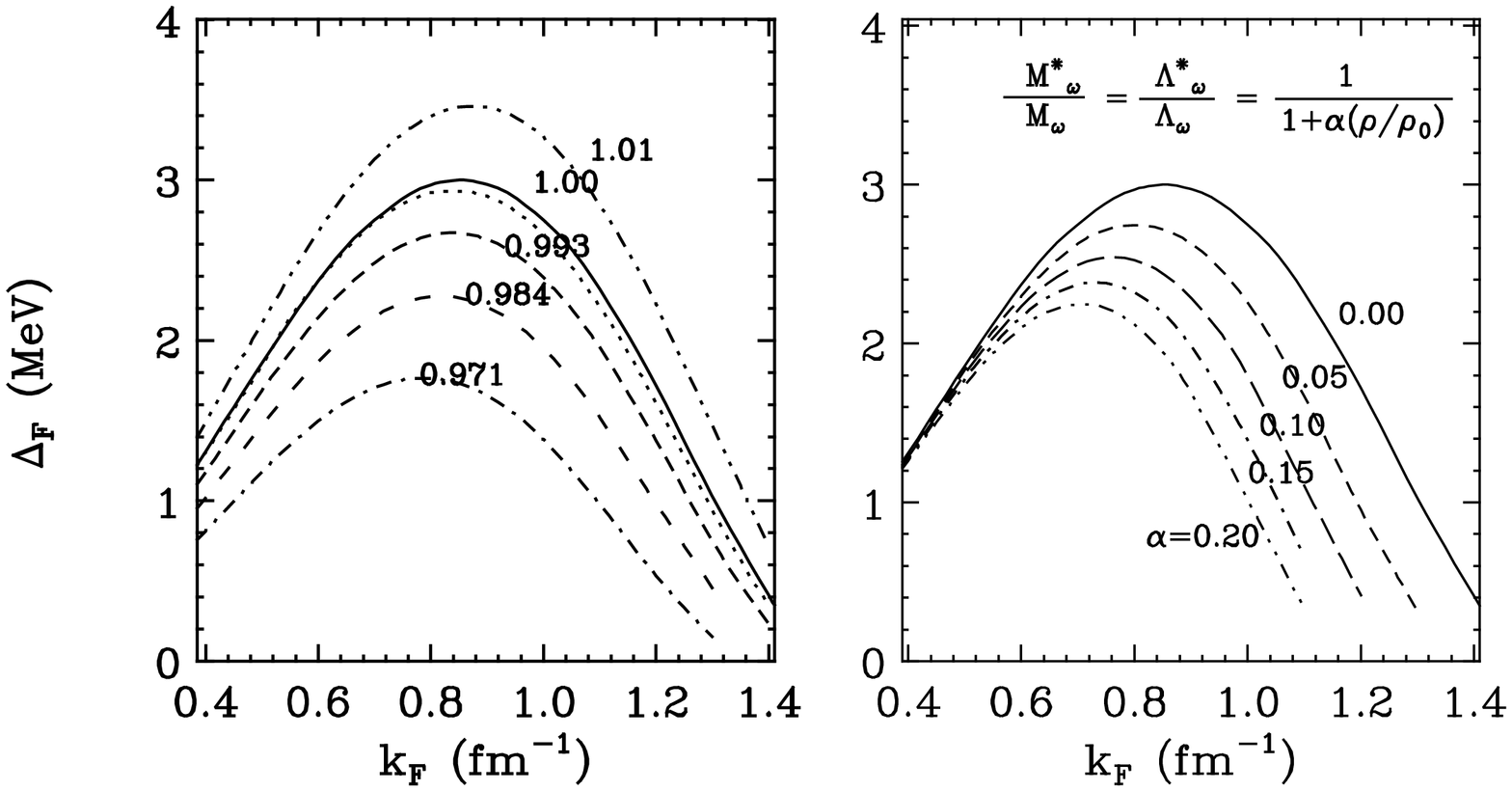}
\caption{
$^1S_0$ pairing gap in neutron matter calculated using the Bonn-B potential
with scaled $\omega$ meson mass and free single-particle energy spectrum.
Left panel: results with different $\omega$ meson mass ratios
$M^*_{\omega}/M_{\omega} = \Lambda^*_{\omega}/\Lambda_{\omega}$, indicated 
near the corresponding curves. 
The solid curve corresponds to unit ratio.
Right panel: results according to Brown-Rho scaling with different
values of the scaling parameter $\alpha$.} 
\label{f:gapmes}
\end{figure}

We have done two sets of calculations:
One is obtained by just changing the meson mass
(and consistently the mass cutoff)
in a range independent of the density;
in the other one the change of the mass is connected to the density
according to the Brown and Rho scaling model \cite{bro}:
\begin{eqnarray}\label{e:2}
 \frac{m^*_{\omega}}{m_{\omega}} =
 \frac{\Lambda^*_{\omega}}{\Lambda_{\omega}} =
 {1 \over 1 + \alpha(\rho/\rho_0)} \:,
\end{eqnarray}
where $\alpha$ is a constant and $\Lambda$ is the mass cutoff.
This scaling of hadron masses has been proposed to partially restore
the chiral symmetry of the QCD Lagrangian in nuclear matter.
In Fig.~\ref{f:gapmes} the results are plotted for the two sets of
calculations.
In order to demonstate the qualitative effect
we solve the gap equation with a free single-particle spectrum.
The Bonn-B potential has been used as pairing interaction
with all meson parameters frozen except for the $\omega$ meson.
On the left side of the figure the first set of results is plotted.
Changing the mass of the $\omega$ meson independently of the neutron density
amounts to the same change of the range of the interaction and therefore a
deviation of the pairing gap from the original curve is found at any density.
It is worth noticing that decreasing the $\omega$ meson mass the peak value
of the energy gap shifts to slightly lower densities, because the average
distance of the competition between long-range and short-range components
of the interaction increases and hence the
pairing peak shifts to a lower density.
On the right side of the figure the deviation of the gap from the original
value is increasing with density simply because the reduction of the
$\omega$ mass also increases.
The effects is stronger for higher scaling parameter.
The shift of the peak value has the same explanation as earlier.

In summary, we have discussed the influence of the short-range part of the
interaction on the $^1S_0$ pairing gap in neutron matter.
Mechanisms which come into play at short distance, such as particle-particle
ladder correlations, three-body forces, and meson masses, have been discussed
in the solution of the BCS gap equation.
These effects appear to point altogether to a reduction of the gap
which seems however small compared to that effected by polarization
corrections to the interaction \cite{rev}.
Unfortunately, at the present time no definite quantitative conclusions
can be drawn due to too many theoretical incertainties.

\section*{Acknowledgments}

We are indebted to Prof. J. W. Clark for drawing our attention to the
study of the three-body force in relation with pairing.
One of us (W.~Z.) would like to thank INFN-LNS and
the Physics Department of the Catania University (Italy)
for their hospitality during the preparation of the present work.
This work has been supported in part by the Chinese Academy of Science
within the one Hundred Person Project,
Knowledge Innovation Project (No. KJCX2-SW-N02),
and the Major State Basic Research Development Program of China under
No.~G2000077400.

\end{document}